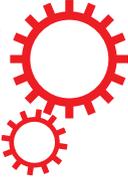

# OPEN  Competing covalent and ionic bonding in Ge-Sb-Te phase change materials

Saikat Mukhopadhyay[1], Jifeng Sun[1,2], Alaska Subedi[3], Theo Siegrist[2] & David J. Singh[4]

$Ge_2Sb_2Te_5$ and related phase change materials are highly unusual in that they can be readily transformed between amorphous and crystalline states using very fast melt, quench, anneal cycles, although the resulting states are extremely long lived at ambient temperature. These states have remarkably different physical properties including very different optical constants in the visible in strong contrast to common glass formers such as silicates or phosphates. This behavior has been described in terms of resonant bonding, but puzzles remain, particularly regarding different physical properties of crystalline and amorphous phases. Here we show that there is a strong competition between ionic and covalent bonding in cubic phase providing a link between the chemical basis of phase change memory property and origins of giant responses of piezoelectric materials ($PbTiO_3$, $BiFeO_3$). This has important consequences for dynamical behavior in particular leading to a simultaneous hardening of acoustic modes and softening of high frequency optic modes in crystalline phase relative to amorphous. This different bonding in amorphous and crystalline phases provides a direct explanation for different physical properties and understanding of the combination of long time stability and rapid switching and may be useful in finding new phase change compositions with superior properties.

$Ge_2Sb_2Te_5$ (GST), a prototype phase-change material (PCM), and in particular its transformation between crystalline and amorphous phases is a subject of ongoing interest because of the complex physics involved and the key role this material plays in phase-change optical (blue-ray) and resistive memories. The material has three solid phases, namely a hexagonal crystalline phase, a cubic crystalline phase (with a locally distorted rhombohedral structure), both conducting, and an amorphous, insulating phase. The cubic and hexagonal phases have long range translational order, while the amorphous phase does not[1–3]. PCM devices use fast transitions between the amorphous and crystalline phases, taking advantage of the remarkably different physical properties of these phases[4]. The origin of this fast switching between these phases[5–11] and the stability[5,6] of these phases has already been discussed in the past. Nevertheless, while the (hexagonal/trigonal) ground state can occur depending on vacancy concentration[1–3,12] and growth conditions, the crystalline phase in most devices is cubic, $c$-GST, with disordered empty lattice sites. This cubic phase ($a$-GST) has a NaCl rocksalt crystal structure, with Te on the anion sublattice and Ge, Sb and empty lattice sites distributed apparently randomly on the cation sublattice. The hexagonal phase can be regarded as an ordering of the cubic structure, and often has a relatively high density of stacking faults that affect properties[13]. The amorphous phase ($a$-GST) exhibits covalent bonding with coordination numbers that are consistent with 8-$N$ rule[14] and smaller Ge-Te and Sb-Te bond lengths compared to its cubic phase[15]. In contrast, the coordination numbers for Ge and Sb are higher in the crystalline phase. This phase has been described in terms of resonant bonding[16], a chemical concept that originates in the description of materials such as benzene and graphite, and has been generalized to the description of covalent systems with high symmetry structures. Other ways of stabilizing high coordinated local structures are metallic bonding, where electron kinetic energy balances Coulomb attraction between electrons and ions, and ionic bonding where the Madelung energy, which favors distortions, balances closed shell repulsions[17]. These are non-directional interactions.

Resonant bonding in benzene and graphite stabilizes multicenter covalent bonds so that structures with high local symmetry form. In the case of graphite, the ground state is described as a mixture of two different $sp^2$ alternating single-double bonding patterns on the six-membered carbon rings, leading to a highly stable high symmetry structure with stronger covalent bonds than the hypothetical lower symmetry structure with distorted rings that would occur with just one of these[16,18]. In graphite and graphene the high symmetry resonantly bonded structure is semi-metallic and conducting in contrast to what is predicted for lower symmetry structures. In the case of GST, the high symmetry crystalline phase has octahedral, six-fold coordinated metal sites and is conducting while the amorphous phase, which may be characterized as distorted from a local point of view is insulating, supporting the analogy[2,16,19,20]. However, given the local lattice distortions noted from both experimental[21,22] and theoretical[3] investigations, it is clear that the meaning of "resonant bonding" in GST is very different from that in graphite, and the precise nature has not been fully established. In the resonant bonding picture discussed by Lucovsky and White[16] and Littlewood[23], covalent bonds stabilize highly symmetric states via sharing of electrons in the bonding network as in the case with the graphene. Naively, this has two consequences: (i) it implies a stabilization of the covalent bonding in crystalline PCMs, and (ii) it suggests that the basic bonding mechanism for amorphous and crystalline GST is the same, e.g. covalent bonding, although the details may differ. Therefore, it does not offer a direct answer to why the physical properties of these phases are so different. Furthermore it should be noted that in materials like graphene, resonant bonding leads to more stable covalent bonding than that in the (hypothetical) distorted structures, with the important result that graphene is a particularly strong material. Here we show that the stabilization of the cubic structure via resonant bonding is more subtle and has a relationship with the covalently bonded non-polar phases of oxide ferroelectrics. This is based on analysis of the electronic structures and bonding of models for *c-GST* and *a-GST*.

[1]Materials Science and Technology Division, Oak Ridge National Laboratory, Oak Ridge, TN 37831-6056 USA. [2]Department of Chemical and Biomedical Engineering, FAMU-FSU College of Engineering, Tallahassee, FL 32310, USA. [3]Max Planck Institute for the Structure and Dynamics of Matter, Hamburg, Germany. [4]Department of Physics and Astronomy, University of Missouri, Columbia, MO 65211-7010 USA. Correspondence and requests for materials should be addressed to S.M. (email: mukhopadhyas@ornl.gov)

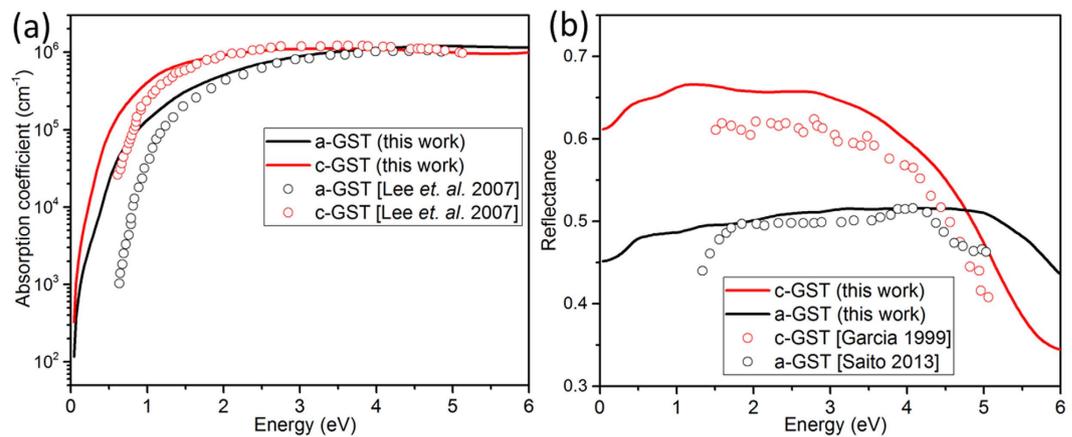

**Figure 1.** (**a**) Absorption coefficient and (**b**) Reflectance of *c-GST* and *a-GST* as a function of energy based on the TB-mBJ functional. The corresponding experimental data are also shown for a better comparison. The experimental data for absorption coefficient was taken from Lee *et al.*[57] and the Reflectance data for *c-GST*[58] and *a-GST* were taken from Garcia *et al.*[58] and Saito *et al.*[59], respectively.

## Results

A special quasirandom structure for *c-GST* and a structure derived using quenched *ab initio* molecular dynamics for *a-GST* were used. The calculated optical absorption coefficient and reflectivity for these two structural models are in accord with experiment over a wide range of energy [Fig. 1] and in particular, the optical reflectivity of the amorphous phase is much lower than the cubic phase in the visible, a property that is exploited in optical storage applications. The absorption coefficient was previously calculated from the extinction coefficient[24] and as well, using first principle calculations with a larger supercell[24,25] and they are in great agreement with our calculated absorption coefficient. Given the optical properties are strongly connected to the electronic structure and bonding, the inference is that these models contain the main differences in bonding that lead to the different optical properties of *c-GST* and *a-GST* and therefore, can be used to understand the differences in electronic structure. It is worth mentioning here that although both *c-GST* and *a-GST* are small gap semiconductors, high reflectivity was noted in the low-energy regime due to the high refractive index and high dielectric constant. The semiconducting nature, however, can be seen in the optical conductivity plot [Fig. 2] where *c-GST* and *a-GST* starts conducting only at energies higher than ~0.2 *eV* and 0.3 *eV*, respectively. These are the calculated band gaps of the respective systems.

A comparison of the electron localization function (ELF)[26] of *c-GST* and *a-GST* [Fig. 3] reveals a difference in the bonding between the two phases. In the amorphous phase, the distribution of the local maxima of the ELF is highly asymmetric about atom centers, and high values of ELF exist along lines connecting pairs of atoms. This implies covalent bonding in *a-GST*, in agreement with previous work[20,27]. We calculated the effective coordination numbers; $N_i = \Sigma_i w_i$, in terms of bond weight of the $i^{th}$ bond; $w_i = \exp\left[1 - \left(\frac{l_i}{l_{av}}\right)^6\right]$, and weighted average bond length; $l_{av} = \frac{\Sigma_i l_i \exp\left[1 - \left(\frac{l_i}{l_{min}}\right)^6\right]}{\exp\left[1 - \left(\frac{l_i}{l_{min}}\right)^6\right]}$, where $l_{min}$ is the smallest bond length of the Ge and Sb centered coordination pol-yhedra[28]. The calculated coordination number [Ge (2.9–4.4) and Sb (2.7–3.5)] are consistent with the 8-*N* rule, which further supports covalent bonding in *a-GST*.

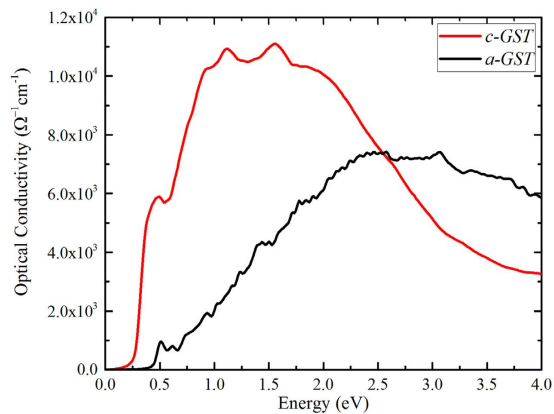

**Figure 2. Optical conductivity of *c-GST* (red) and *a-GST* (black).** We plot the direction average of σ, i.e., $\sigma = \frac{(\sigma_{xx} + \sigma_{yy} + \sigma_{zz})}{3}$ to eliminate artificial anisotropy.

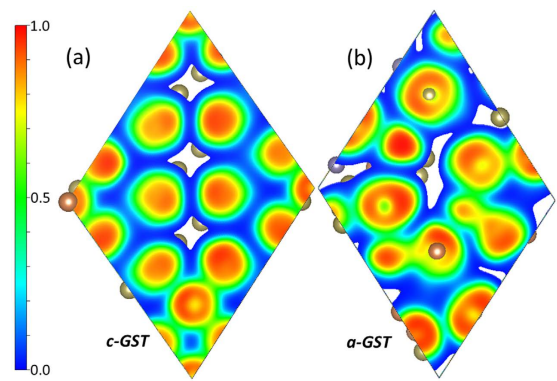

**Figure 3.** Electron localization function of (**a**) *c-GST* and (**b**) *a-GST* at an iso-surface value of 0.02 e/Å³. The violet, orange, and brown spheres stand for Ge, Sb and Te atoms, respectively.

On the other hand, the distribution of the local maxima of the ELF in *c-GST* is more symmetric around the ions, i.e. consistent with the structural motivation for the resonant bonding in *c-GST*. However, local maxima in the ELF are not evident on lines connecting pairs of atoms in the relaxed structure. This would indicate reduced covalent sharing of electrons in *c-GST* on a local level. In a simple resonant bonding picture, high values of ELF along the lines joining pairs of atoms is expected as it would have enhanced covalency due to the resonant characteristic. However, the ELF of *c-GST* as shown in Fig. 3(a) lacks this feature.

Both the *a-GST* and *c-GST* models show strong local structural distortions. The metal atoms (Ge and Sb) in *c-GST* are displaced with respect to their ideal rocksalt positions in the relaxed structure (Fig. S1 in Supplementary Information), and the Te-Sb-Te, Te-Ge-Te and Ge-Te-Sb bond-angles vary widely, ranging from 164⁰ to 175⁰. This is in contrast to tetradymite, $Bi_2Te_3$[29] and also $Sb_2Te_3$[30] both of which exhibit less variance than *c-GST* (note that $Sb_2Te_3$ is a low temperature phase change material[31]).

The nearest neighbor distances of Ge and Sb atoms in *c-GST* and *a-GST* are given in Fig. 4. A number of studies have focused on the average characteristic bond lengths[3,32], and it has been shown from both the theoretical and experimental studies that the Ge-Te bond lengths [DFT: 2.78[32], XRD: 2.61[15], EXAFS: 2.63[33]] are slightly shorter than Sb-Te bonds [DFT: 2.93[32], XRD: 2.85[15], EXAFS: 2.83[33]] in *a-GST*. These bonds were reported to be elongated in *c-GST* [Ge-Te: 2.83 and Sb-Te: 2.91] from XRD studies[15]. We also find that the average Sb-Te bond length [3.11 Å for *c-GST* and 3.09 Å for *a-GST*] is longer than the average Ge-Te bond length [3.01 Å for *c-GST* and 2.87 Å for *a-GST*] in both the phases, and the average Sb-Te bond length is longer in *c-GST* compared to *a-GST*. The longer average bond length in *c-GST* again suggests reduced covalency in this phase.

It should be noted that the distribution of bond lengths is narrower in *c-GST* than *a-GST*. As a result, *c-GST* has fewer long bonds than *a-GST* (again see Fig. 4) even though the average bond lengths are larger in *c-GST*. This is a consequence of the lower average coordination numbers in *a-GST* (note that *a-GST* has a larger volume, which partly compensates this).

The lengthening of the Ge-Te bonds will strongly affect the vibrational properties of *c-GST*. This is reflected in the vibrational properties, which can be measured by inelastic neutron or x-ray scattering. Figure 5 shows the calculated phonon density of states for *c-GST* and *a-GST* projected on the constituent atoms. The results are qualitatively consistent with neutron measurements for $GeSb_2Te_4$, which is a related phase change material[34]. The most significant difference is the simultaneous hardening of acoustic modes and softening of higher frequency optic modes in *c-GST* with respect to *a-GST*. While the former acts to enhance the phonon group velocity, the later reduces the phase space for three-phonon scattering processes yielding higher thermal conductivity in *c-GST*. This is consistent with the measured thermal conductivity behavior in these phases[12,35]. Although the characteristic Sb peaks in Fig. 5 do not undergo dramatic changes, the modes around 3 THz are suppressed in *a-GST* compared to that in *c-GST*.

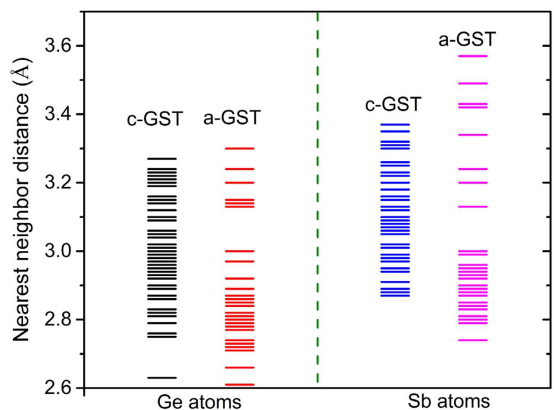

**Figure 4. Nearest neighbor distance of Ge and Sb atoms in *c-GST* and *a-GST*.**

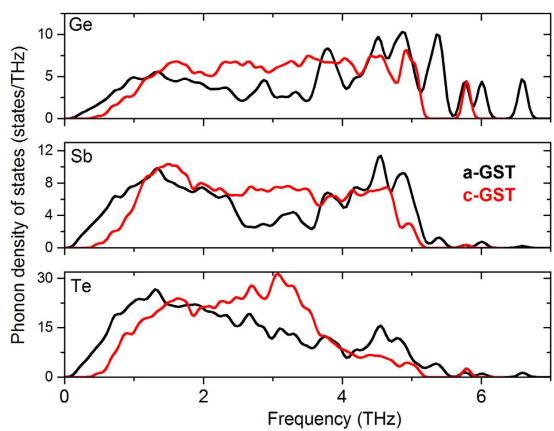

**Figure 5. Phonon density of states projected on constituent atoms of *c-GST* (red line) and *a-GST* (black line) phase.**

The hardening of acoustic modes in *c-GST* can be traced to the disappearance of the very long Ge-Te and Sb-Te bonds in this phase, and possibly the alignment of the bonding orbitals. This leads to a shift of their characteristic peak frequency to a higher frequency (at ~1.5 THz). It is interesting to note that the hard modes containing significant Te character around 5 THz in *a-GST* are suppressed in *c-GST* and new softer modes appear around 3 THz. These modes were previously assigned to the transverse and longitudinal optic modes in $\alpha$-GeTe[34] but we find that these modes are more complex in nature and involve vibrational motions of Sb and Ge cages and may reflect some degree of Te-Te bonding. However, the most striking effect of crystallization is the removal of pure Ge stretching modes at 6 and 6.6 THz resulting in the optic mode softening in *c-GST*. As seen in Fig. 5, there is only a small contribution from Te and Sb in this frequency range. This can be attributed to the elongation of the shortest Ge-Te bonds in *a-GST* upon crystallization as shown in Fig. 4. Optic mode softening upon crystallization was observed previously[34] but was discussed in terms of Sb-Te bonds, while we find that the Ge-Te bonds play a key role. The softening of the high frequency phonons in *c-GST*, which is related to the weakening of nearest neighbor bonding, again shows that the bonding in *c-GST* is less covalent and, rather more ionic in nature, as compared to *a-GST*.

The difference in bonding in *c-GST* and *a-GST* can also be understood in terms of the changes in cross-gap hybridization. The atom- and orbital-resolved densities of states of *c-GST* and *a-GST* are shown in Fig. 6. The Ge and Sb *s states* are energetically separated from the corresponding *p* states and are fully occupied. The Ge and Sb *p* states are higher in energy and take part in bonding. In both phases, it is apparent that the nature of conduction bands is quite different from the valence bands. More specifically, the valence bands have much higher Te *p* character than the conduction bands and conversely, the conduction bands have more Sb and Ge *p* character than the valence bands. This reflects the polar nature of the bonds and is an important reason for considering whether an ionic starting point can be used for understanding the bonding of GST. Besides this we note that there is less Te *p* character above the gap in *c*-GST, which implies that the bonding is more polar and, consequently, less covalent in this phase. This can also be discussed from the point of view of an ionic system with cross-gap hybridization (i.e. hybridization of nominally occupied Te *p* states and nominally unoccupied Ge/Sb *p* states). The starting point is then an ionic picture in which the chalcogen *p*-states are occupied and the metal *p* states are unoccupied, i.e. forming the valence and conduction bands, respectively. The metal-chalcogen hybridization is then cross-gap hybridization as has been discussed for GeTe[36], which has a polar ferroelectric type structure. We use this as a starting point for discussing the bonding. A comparison of the total density of states [Fig. S2 in Supplementary Information] in these phases shows an increase in splitting of the valence and conduction bands in going from *c*-GST to *a*-GST, not just at the band edges, but over the whole valence energy region. Since the valence band contains the nominally bonding states, and the conduction bands antibonding states, this opening suggests stronger covalency in *a*-GST. In any case, this change is reflected in the optical properties. In particular, the optical properties [Fig. 1] show changes not only at energies comparable to the band gap, which is in the infrared below 1 eV, but at higher energies into the blue part of the visible and beyond. Furthermore, the mixing amount of metal *p*-character in the valence bands, especially the lower valence bands is higher in *a*-GST. This is very similar to what is seen in PbTiO$_3$[37]. In oxide ferroelectric materials[37–39], this cross-band gap hybridization results in enhanced Born-effective charges (Z*) as compared to the nominal valence charges on the individual atoms and ultimately is responsible for the ferroelectricity. In other words, in oxide ferroelectrics the hybridization of the nominally occupied O *2p* states with nominally unoccupied states on the cations e.g. the Ti *3d* states of BaTiO$_3$ and the Pb *6p* and Ti *3d* states in PbTiO$_3$ leads to enhanced Born charges that underlie the ferroelectricity of these materials. As mentioned, high symmetry ionic structures such as perovskite and rocksalt can occur because of an interplay of closed shell ionic repulsions and the Coulomb (Madelung) attraction of opposite sign charged ions. While the Coulomb interactions would favor distortions that would have low symmetry structures with some short bond lengths, the stronger functional dependence on bond length of the repulsive interaction forces ions to take the highly coordinated structures characteristic of simple ionic compounds. Cross gap hybridization of metal *p* and *d* states with oxygen *p* states enhances the tendency towards distortion and leads to ferroelectric instabilities in materials such as BaTiO$_3$ and PbTiO$_3$. These enhanced Born charges, while reflecting cross gap covalency, can be viewed as an enhancement of the nearest neighbor Coulomb attractions.

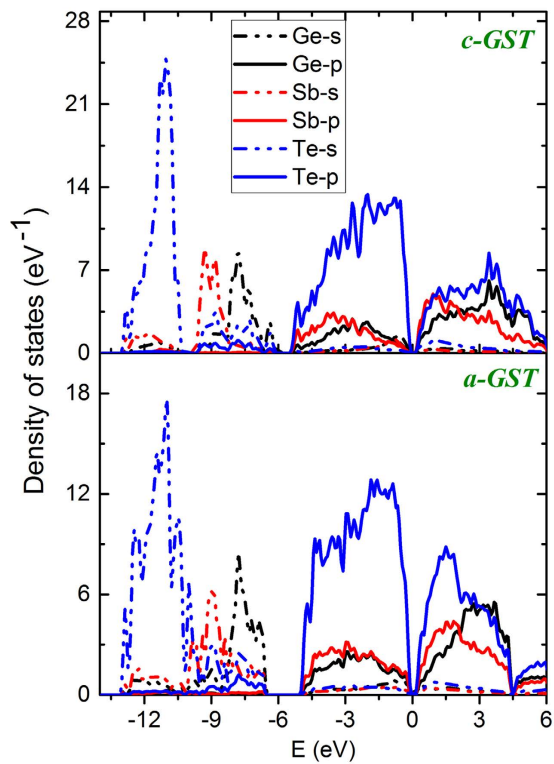

**Figure 6. Orbital resolved density of states of *c*-GST (top panel) and *a*-GST (bottom panel) projected on Ge, Sb and Te atoms.**

We calculated the Born effective charges on the individual atoms in *c*-GST and *a*-GST using density functional perturbation theory. Due to the lack of symmetry in these phases, neither the dielectric tensors nor the Born effective charges have any symmetry. For a better comparison, we averaged the diagonal terms of dielectric constant $[[\varepsilon_{ij}^\infty]/3]$ and Born effective charge $[Tr[Z_{ij}^\infty]/3]$ tensors [Fig. S3 in Supplementary Information].

Our calculations overestimate the dielectric constant ($\varepsilon^\infty$) for *c*-GST (42.7) and *a*-GST (22.9) relative to the measured values of 33.3 and 16.0, respectively[2]. This is commonly the case in density functional calculations and is related to the underestimation of band gaps.

| Born effective charges (e) | Ge | Sb | Te |
|---|---|---|---|
| c-GST | 4.89 | 7.34 | −4.44 |
| a-GST | 2.19 | 2.02 | −1.67 |

**Table 1. Average Born effective charges on individual atoms of *c-GST* and *a-GST*.**

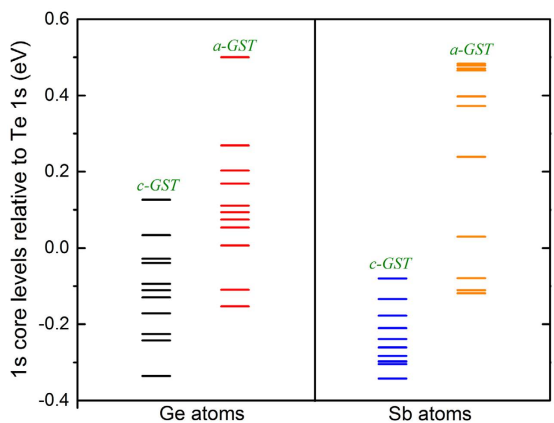

**Figure 7. Relative shift of the 1s core levels of Ge and Sb in *c-GST* and *a-GST* with respect to their corresponding average Te 1s core level.** The energy zero is set so that the average of all the Ge or Sb 1s levels including both models is zero.

Nevertheless, the magnitudes of the average Born effective charges (Table 1) in *a-GST* are notably smaller implying that charges are delocalized along the bonds, a true signature of covalently bonded system. On the other hand, the Born charges are enhanced in *c-GST*, a signature of a situations like in $ABO_3$[38–41] where one has an essentially ionic bonded material with Born charges enhanced by cross gap hybridization. This is a different situation from ordinary polar-covalent materials, such as GaAs, where the Born charges are reduced from the nominal valences by hybridization. In any case, large Born charges soften the polar transverse optical modes in a material and make the lattice susceptible to polar distortions, which then leads to high dielectric constant, as seen in GST. In general high dielectric constants screen defects and favor high mobility in semiconductors[42], and this is expected to be important for the electronic properties of *c-GST*, which has significant conductivity in spite of a very strongly disordered cation lattice.

The core level shifts are also instructive. Figure 7 shows the relative positions of the *1s* core levels of Ge and Sb relative to the average *1s* position core level of Te. As seen in Fig. 7, the *c-GST* cation *1s* levels are lower with respect to *a-GST* by several tenths eV. In ionic materials, stable structures are ones where the cations occur on sites that are favorable for positively charged species, in other words, sites that are repulsive to electrons, and vice versa for anions. We carried out calculations for $PbTiO_3$ with the PBE GGA functional. We find that the deep core Pb *1s* and Ti *1s* levels shift upward in energy relative to the average O *1s* position in going from the higher symmetry cubic structure to the tetragonal ferroelectric structure, similar to the behavior of GST going from the cubic to the amorphous phase (note that the *1s* level position is a measure of the on-site Coulomb potential). This shift is by 0.45 eV and 0.52 eV for Pb and Ti, respectively, in $PbTiO_3$. Thus as measured by the core levels, which follow the Coulomb potential, the cation sites are more favorable from an ionic point of view in the distorted structure even though the distorted structure is more covalent, consistent with the above discussion. We find the same situation in the core level positions of GST in going from *c-GST* to *a-GST*. *c-GST* is a disordered material with a net cubic symmetry that precludes an overall net polarization similar to relaxor ferroelectric oxides. The large Born charges, which come from cross gap hybridization, again similar to the relaxors, imply local polar behavior as polarization is given by the product of the atomic displacement from the ideal symmetric site times the Born charge.

## Discussion

In summary, we investigated models of *c-GST* and *a-GST* and show a different bonding nature between the two phases, specifically a greater degree of covalency in *a-GST*. The results show a connection between this phase change material and classical ferroelectric materials, such as $BaTiO_3$ and $PbTiO_3$. This provides additional richness to the concept of resonant bonding and in particular shows new aspects beyond the classical resonant bonding of materials like graphite. There are also certain consequences. First of all, switching between an ionic and a more covalent structure does not require diffusion of atoms and therefore can be fast. However, it does require coordinated motion of many atoms because coordination is key to a covalent structure, as expressed in the 8-N rule, while Coulomb interactions with extended range are important for ionic structures. This characteristic provides a framework in which one can have both fast crystallization at elevated temperature and an extremely long lived metastable amorphous phase at lower temperature.

Secondly, the cubic phases of classic oxide ferroelectrics, specifically $KNbO_3$ and $BaTiO_3$ show highly anisotropic long correlations of atomic displacements along chains of atoms[43,44]. These are a consequence of the directional covalent interactions and covalent bond competition, a concept which explains many features of the structure of complex ferroelectrics and can be used to construct precise force fields for simulating properties[45,46]. This suggests experiments on cubic GST looking for long range correlations of atomic displacements associated with the *p*-electron bonding network. In any case, the present results show an interplay of ionic and covalent bonding in GST, which is important in understanding the physical properties of this material and provides additional richness to the concept of resonant bonding.

## Methods

The internal coordinates were optimized starting with the experimental lattice parameter[47] using projector augmented wave (PAW)[48,49] approach within the *Generalized Gradient Approximation* (*GGA*) of density functional theory. We used the Perdue-Burke-Ernzerhof (PBE)[50,51] exchange-correlation functional as implemented in VASP[52,53]. A $2 \times 2 \times 2$ Monkhorst-Pack grid was used for K-points sampling of the Brillouin zone of the unit cell with an energy cutoff of 550 eV. The convergence threshold for the energy was set to $10^{-8}$ *eV* and $10^{-6}$ eV/Å for its gradient. We employed finite displacement method using (VASP-phonopy[54]) using periodic boundary conditions to calculate the harmonic force constants for all the phases discussed in this study with a higher convergence threshold ($10^{-8}$ eV/Å) for energy. For *c-GST*, we generated two special quasi-random structures (SQS). These were a 27-atom and a 45 atom cells. We used the prescription of Zunger *et al.*[55] and a Monte Carlo optimization as implemented in the ATAT code. The SQSs were generated such that all the pair correlation functions (PCF) up to the next nearest neighbors are identical to the average PCF of an infinite random structure [see Fig. S4 in Supplementary Information]. The results were similar for the two cells, and here we focus on the larger one. For *a-GST* we used the larger SQS structure, increased the volume by 9.5% corresponding to the volume difference of the phases, raised the temperature to 2000 K in 2 ps, held at 2000 K for a further 3 ps and then cooled to 300 K over 1 ps in an ab initio molecular dynamics run. We then fully relaxed the atomic positions. The optical properties were calculated using the linearized augmented planewave method as implemented in the WIEN2k[56] code. Spin-orbit coupling was included. The self-consistent calculations were performed with a $4 \times 4 \times 4$ k-point grid of 36 k points in the irreducible Brillouin zone. Optical properties were calculated using a $8 \times 8 \times 8$ k-mesh. We used well converged basis sets with LAPW basis size corresponding to $R_{MT}K_{max} = 9.0$, where $R_{MT}$ is the smallest sphere radius and $K_{max}$ is the plane-wave cutoff parameter. LAPW sphere radii of 2.47 Bohr, 2.5 Bohr and 2.5 Bohr were used for Ge, Sb and Te, respectively.

### Acknowledgements

SM gratefully acknowledges support from the U. S. Department of Energy, Office of Science, Office of Basic Energy Sciences, Materials Sciences and Engineering Division. This work was partially supported as part of S3TEC an Energy Frontier Research Center funded by the Department of Energy, Office of Science, Basic Energy Sciences under award #DE-SC0001299/DE-FG02-09ER46577(DJS). J.S. acknowledges a graduate student fellowship, funded by the Department of Energy, Basic Energy Science, Materials Sciences and Engineering Division through the ORNL GO! program. SM thankfully acknowledges computing resources from Research Services at Boston College and Texas Advanced Computing Center (TACC) at The University of Texas at Austin. Helpful discussions with Dr. Mao-Hua Du are greatly acknowledged. T.S. acknowledges funding by the Department of Energy, Office of Science, Basic Energy Sciences under award #DE-SC0008832.

### Author Contributions

D.J.S. conceived the idea, S.M., J.S., D.J.S. and A.S. conducted the calculations, S.M., D.J.S. and T.S. analyzed the results. S.M. and D.J.S. wrote the manuscript. All authors reviewed the manuscript.

### Additional Information

**Supplementary information** accompanies this paper at http://www.nature.com/srep

**Competing financial interests:** The authors declare no competing financial interests.

**How to cite this article**: Mukhopadhyay, S. *et al.* Competing covalent and ionic bonding in Ge-Sb-Te phase change materials. *Sci. Rep.* **6,** 25981; doi: 10.1038/srep25981 (2016).